\pdfoutput=1
\documentclass[aps, prc, 12pt,preprint,tightenlines,nofootinbib]{revtex4}

\usepackage[toc,page]{appendix}
\usepackage{appendix}
\usepackage[caption=false]{subfig}
\usepackage{epsfig}
\usepackage{hyperref}
\usepackage{amssymb}
\usepackage{color}
\usepackage{slashed}
\usepackage{amsmath}
\newcommand{\be}{\begin{equation}}
\newcommand{\ee}{\end{equation}}
\newcommand{\ba}{\begin{eqnarray}}
\newcommand{\ea}{\end{eqnarray}}
\newcommand{\bd}{\begin{displaymath}}
\newcommand{\ed}{\end{displaymath}}

\newcommand{\commentout}[1]{{}}

\begin{document}

\title{Electromagnetic recombination spectra at the quark-hadron phase transition}
\author{Clint Young}
\email{youngc@nscl.msu.edu}
\affiliation{Department of Physics and Astronomy and National Superconducting Cyclotron Laboratory, Michigan State University, East Lansing, MI 48824, USA}
\author{Scott Pratt}
\email{pratt@nscl.msu.edu}
\affiliation{Department of Physics and Astronomy and National Superconducting Cyclotron Laboratory, Michigan State University, East Lansing, MI 48824, USA}
\date{\today}

\begin{abstract}
When quarks hadronize, they accelerate. Because they carry electric charge, they must radiate light as they accelerate and hadronize. This is true not only in jets but also in heavy ion collisions,
where a thermalized plasma of quarks and gluons cools into a gas of hadrons. First, direct emission of photons from two quarks coalescing into pions is calculated using
the quark-meson model. The yield of final-state photons to pions is found to be about $e^2/g^2_{\pi qq}$, which is on the order of a percent. Second, the yield of photons from the decay of highly
excited color singlets, which may exist ephemerally during hadronizaton, is estimated. Because these contributions occur late in the reaction, they should carry significant elliptic flow, which may
help explain the large observed flow of direct photons at RHIC by the PHENIX Collaboration at the Relativistic Heavy Ion Collider (RHIC). The enhanced emission also helps explain PHENIX's
surprisingly large observed $\gamma/\pi$ ratio.
\end{abstract}

\maketitle

\section{Introduction}

Electromagnetic radiation is produced from multiple sources in heavy ion collisions. Of particular interest is the radiation produced during the deconfined phase by the scattering and annihilation of 
quarks. This thermal radiation is not that of a blackbody yet nevertheless provides a reasonable measurement of the temperature reached in these collisions \cite{Kapusta:1991qp, Dion:2011vd}. The 
elliptic flow of the radiated photons gives some measure of when these photons were created: the elliptic flow during the plasma phase starts small and then builds up to the measured values, and the 
photons produced at the various times have flow similar to that of the matter from which they were radiated. The first predictions of thermal photon production had relatively small elliptic flow, thanks 
to a large component of the photons coming from the earliest times in the collision. However, measurements by the PHENIX collaboration at the Relativistic Heavy Ion Collider show $v_2(p_T)$ of photons 
approaching that of hadrons \cite{Petti:2013iza}, which suggests underestimation of photons emitted later in the reaction. PHENIX's observed yield of direct photons is also larger than what was first 
predicted using thermal rates of production \cite{Dion:2011vd}. Recently, considerable work both in examining thermal photon rates and of viscous corrections has quantified both the strong agreement 
between theory and experiment for much of the momentum range for photon production, while showing exactly where novel mechanisms for photon production might be at play \cite{Paquet:2015lta}.

Possible explanations for the large elliptic flow include initial flow, incomplete thermalization in the quark sector at early times, and the enhancement of quark degrees of freedom in the Polyakov 
Nambu-Jona-Lasinio and the ``semi-QGP'' models \cite{Ratti:2005jh, Kashiwa:2013gla, Hidaka:2015ima}. In this paper, we consider a neglected source of photons sure to exist: the production of 
photons at the point of 
recombination of quarks and gluons into hadrons. This source is analogous to the recombination spectra studied in plasma physics and in cosmology, but it is also similar the the ``fragmentation photons'' 
produced in event generators such as \textsc{pythia} as quarks evolve from large to small $Q^2$. The production of photons at the point of hadronization was also considered by Campbell 
\cite{Campbell:2015jga}; this current work differs from Campbell's in that we are primarily interested in how the electromagnetically charged quarks and anti-quarks radiate as they hadronize, as 
opposed to how gluonic degrees of freedom might create light as they disappear.

\begin{figure}
\centerline{\includegraphics[width=0.7\textwidth]{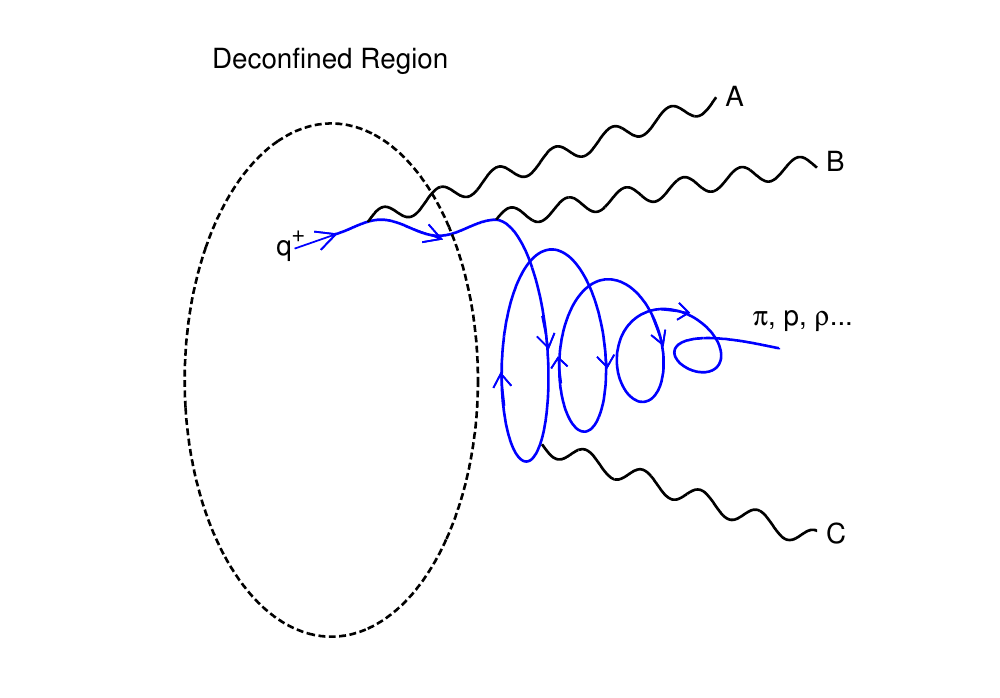}}
\caption{\label{fig:curly}
(Color online) An illustration of different sources of photons. Emission from the plasma stage (A) is induced by collisions and annihilations of quarks. In (B), the transition to a bound state is electromagnetic, leading
 to photon emission. If loose color singlets are first formed, photons are also emitted during the spontaneous emission of excited states (C).}
\end{figure}

Figure \ref{fig:curly} illustrates three contributions to direct photon production from deconfined quarks. Emission occurs as unbound quark states are jostled in the medium, as is well known. But 
a significant jerk occurs when the quarks begin to form bound states, leading to electromagnetic radiation. Finally, some of these bound states are initially excited, and must undergo transitions 
(sometimes electromagnetic) to ground states. This cartoon emphasizes how large the acceleration of quarks is at the end of the heavy ion collision, but work in quantifying this is necessary.

\begin{figure*}
\subfloat[\label{fig:photonsVspiplus}]{%
  \includegraphics[trim=1cm 2cm 11cm 20cm, clip=true, width=.49\linewidth]{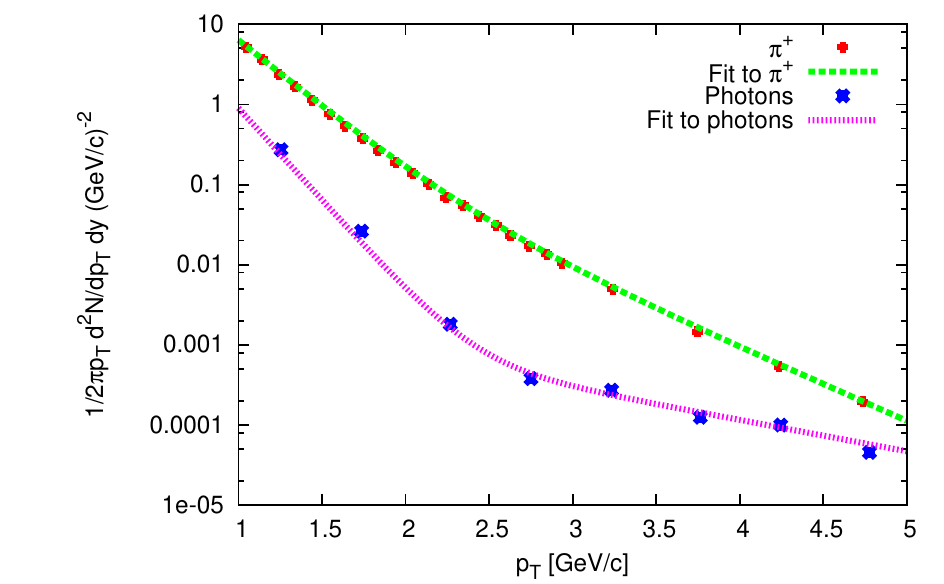}%
}
\hfill
\subfloat[\label{fig:ratio10to20}]{%
  \includegraphics[trim=1cm 2cm 11cm 20cm, clip=true, width=.49\linewidth]{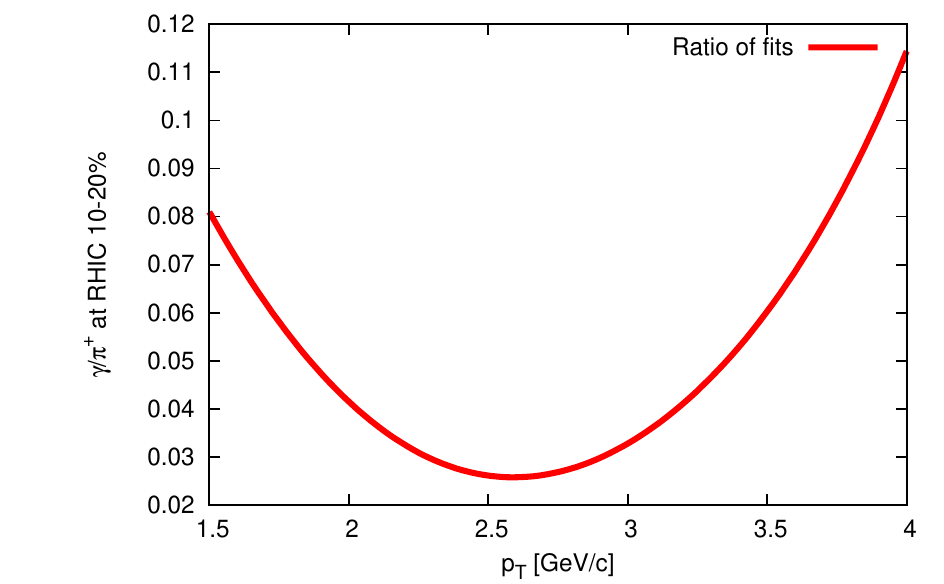}%
}
\caption{(Color online) a.) The yields of $\pi^+$ and direct photons in the 10-20\% centrality class of Au+Au $\sqrt{s}=200\;{\rm GeV}$ collisions from the PHENIX Collaboration 
  \cite{PHENIXURL, Adler:2005ig}, together with fits to these yields. b.) the ratio of the yields in Fig. \ref{fig:photonsVspiplus}.}
\label{fig:phenix}
\end{figure*}

Figure \ref{fig:phenix} displays the direct photon spectra as measured by PHENIX \cite{Adler:2005ig} along with the ratio of the direct photon spectra to that of positive pions \cite{PHENIXURL}. 
If a contribution to the photons is to 
explain the large elliptic flow, it must come after the first few fm$/c$ of the collision so that elliptic flow will have built up, and to explain the measured spectra the contribution must be of the 
order of 2\% of the final number of final-state pions. By considering emission related to hadronization, which should occur for times $3-6$ fm/$c$ into the collision, the first criterion is met. For the 
remainder of this paper, we will focus on seeing whether contributions from emission at hadronization could have sufficient strength to give a ratio of the yields of direct photons to $\pi^+$ of 
roughly 2\%.

The unifying theme of this current work is that the transition from unbound to bound states and ultimately to ground states represents a significant acceleration of electrically charged 
particles and therefore an important source of electromagnetic radiation in heavy ion collisions. This paper is broken up as follows: in Section \ref{sec:quark-meson} we use the 
quark-meson model at leading order to describe electromagnetic 
radiation from quarks coalescing directly into pions. This model is similar to Polyakov loop-inspired models in that the 
gluonic degrees of freedom are suppressed. The diagram of interest involves an incoming quark-antiquark pair evolving into a final-state pion and a photon, $q\bar{q}\rightarrow \gamma\pi$. 
Because the outgoing state has only two particles, the photon can carry approximately half the center of mass energy which suggests the process might be a good candidate for photons in the GeV range. 
In Section \ref{sec:l>0} we examine whether the transition to hadrons might lead to yields of excited states beyond the thermal expectation values, using a simple scalar potential model to confine 
the quarks into bound states and two different models for how this transition might occur. Finally, in Section \ref{sec:EM}, we calculate the electromagnetic transitions in these models, 
to determine rates for photon production both from a thermal gas of bound quarks as well as a gas with elevated populations of excited states. In Section \ref{sec:Conclusions}, we summarize 
the main points of the paper as well as suggest where a great deal of future work is necessary.

\section{Photon production in the quark-meson model}
\label{sec:quark-meson}

In the quark-meson model (which ultimately evolved from descriptions 
of meson-nucleon couplings in \cite{GellMann:1960np}), both quarks and mesons are treated as fundamental point-like degrees of freedom, and gluonic degrees of freedom are ignored. The Lagrangian, 
without the electromagnetic coupling, is of the form
\begin{equation}
{\cal L} = \bar{\psi}(i\slashed{\partial}-ig_{qq\pi}{\bf \tau}\cdot{\bf \phi}\gamma_5-m)\psi + {\textstyle \frac{1}{2}}|\partial_\mu \phi|^2 - {\textstyle \frac{1}{2}}m^2_i|\phi_i|^2-V({\bf \phi}){\rm .}
\nonumber
\end{equation}
Here, $V({\bf \phi})$ is the potential which remains after spontaneously breaking the original symmetry in the mesonic sector and explicit chiral symmetry breaking gives masses $m_i$ to the Goldstone 
bosons of the mesonic sector's symmetry breaking. For our purposes, $V({\bf \phi})$ is ignored: we will be concerned with tree-level diagrams where quarks and anti-quarks annihilate into color-neutral 
particles.

When plasmas of quarks at temperatures near 180 MeV cool, they eventually must combine to form the color singlets observed at low temperatures. The quark-meson model can describe this process at these 
high but not asymptotically high temperatures, because one needs to be in the transition range where both hadrons and quarks coexist.
For simplicity, we consider only those diagrams where the outgoing hadrons are pions. Later, we will discuss how rates might grow if additional hadronic states
are included. Additionally, we consider only $2\rightarrow 2$ diagrams. The neglected $2\rightarrow 3$ processess produce lower energy photons, and should not contribute as significantly to the 
$\gtrsim 1$ GeV direct photons measured experimentally. 

At tree level this leaves only 5 leading-order channels where the outgoing states are pions or photons: 
$q\bar{q}\to\pi^0 \gamma$, $q\bar{q}\to\pi^{\pm} \gamma$, $q\bar{q}\to\pi^0 \pi^0$, $q\bar{q}\to\pi^+ \pi^-$, and $q\bar{q}\to\pi^{\pm} \pi^0$.
At the point of hadronization, there exist two processes of the ones above which lead to the production of photons. If these were the only two processes, then our work would be finished: we would only 
have to count the numbers of quarks to determine the number of photons produced. However, there are two competing (indeed, significantly 
more frequent) processes above which lead only to the production of pions. 

\begin{figure*}
\subfloat[\label{fig:PionPhoton}]{
  \includegraphics[trim=2cm 16cm 2cm 3.75cm, clip=true, width=0.45\textwidth]{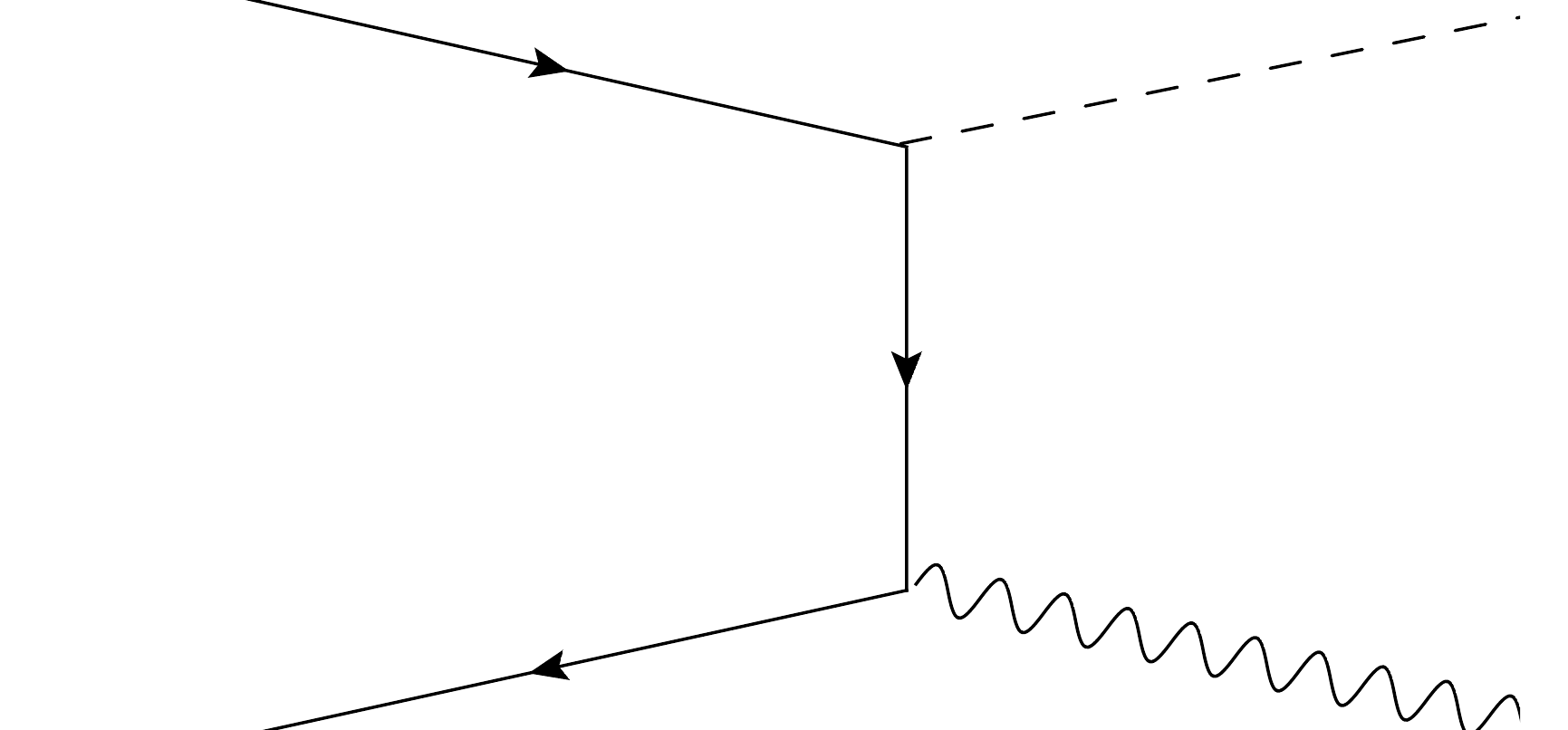}
}
\hfill
\subfloat[\label{fig:PionPion}]{
  \includegraphics[trim=2cm 17cm 2cm 1cm, clip=true, width=0.45\textwidth]{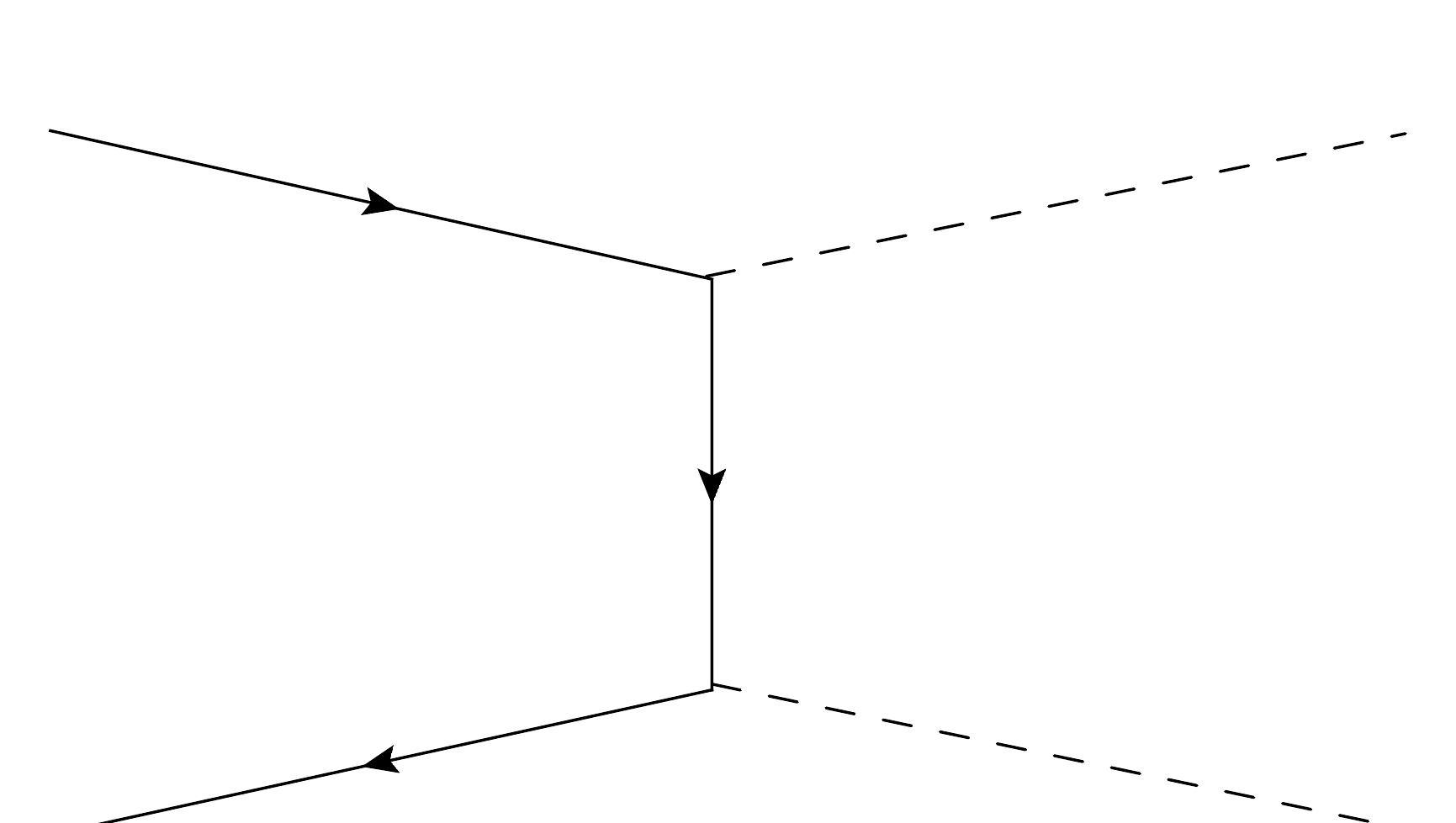}
}
\caption{a.) An example of a diagram contributing to the process of quarks annihilating into a pion and a photon. b.) A 
diagram describing the annihilation of quarks into two pions that competes with the photon production in Fig. \ref{fig:PionPhoton}.
}
\end{figure*}

This situation is encountered frequently in atomic physics, where excited states de-excitate radiatively at 
a rate $\Gamma_{rad}$ as well as de-excitate collisionally at a rate $\Gamma_{coll}$. The {\it quantum efficiency}
\begin{equation}
Q = \frac{\Gamma_{rad}}{\Gamma_{rad}+\Gamma_{coll}}
\end{equation}
gives the fraction of states which de-excitate radiatively; equivalently, if there are $N$ excited atomic states in a given gas, and these excited states only decay once, then there are $QN$ 
photons radiated once all excited states decay.
We are interested in the same quantity, the quantum efficiency of photon production at hadronization.

The first step in calculating this efficiency is to determine the matrix elements for photon production at hadronization. First, we consider $q\bar{q}\to \pi^0 \gamma$: the quarks are coupled to the 
mesons through the $\bar{\psi}{\bf \tau}\cdot{\bf \pi}\psi$ in the quark-meson model. The spin-summed matrix element squared is
\begin{eqnarray}
\sum|{\cal M}_{\gamma \pi^0}|^2=1/2(5e^2/9) g_{q\bar{q}\pi}^2 \sum_{{\rm spins}} \bigg[\bar{v}_\sigma (p^\prime) \gamma_5 \frac{\slashed{p}-\slashed{k}+m}{(p-k)^2-m^2+i\epsilon}\gamma^\mu u_\lambda(p) 
\bar{u}_\lambda (p) \gamma_\mu
\frac{\slashed{p}-\slashed{k}+m}{(p-k)^2-m^2+i\epsilon}\gamma_5 v_\sigma(p^\prime) \nonumber \\
+\bar{v}_\sigma (p^\prime) \gamma^\mu \frac{\slashed{p}^\prime-\slashed{k}+m}{(p^\prime-k)^2-m^2+i\epsilon}\gamma_5 u_\lambda(p) \bar{u}_\lambda (p) \gamma_\mu
\frac{\slashed{p}-\slashed{k}+m}{(p-k)^2-m^2+i\epsilon}\gamma_5 v_\sigma(p^\prime) \nonumber \\
+ \bar{v}_\sigma (p^\prime) \gamma_5 \frac{\slashed{p}-\slashed{k}+m}{(p-k)^2-m^2+i\epsilon}\gamma^\mu u_\lambda(p) \bar{u}_\lambda (p) \gamma_5
\frac{\slashed{p}^\prime-\slashed{k}+m}{(p^\prime-k)^2-m^2+i\epsilon}\gamma_\mu v_\sigma(p^\prime) \nonumber \\
+ \bar{v}_\sigma (p^\prime) \gamma^\mu \frac{\slashed{p}^\prime-\slashed{k}+m}{(p^\prime-k)^2-m^2+i\epsilon} \gamma_5 u_\lambda(p) \bar{u}_\lambda (p) \gamma_5
\frac{\slashed{p}^\prime-\slashed{k}+m}{(p^\prime-k)^2-m^2+i\epsilon}\gamma_\mu v_\sigma(p^\prime) \bigg] \nonumber \\
\nonumber
\end{eqnarray}
\begin{eqnarray}
= 1/2(5e^2/9)g_{q\bar{q}\pi}^2 \bigg\{ 16[p\cdot k (p^\prime \cdot k) -m^2p\cdot p^\prime + m^2(p\cdot k + p^\prime \cdot k) -m^4]/(4(p\cdot k)^2) \nonumber \\
+2\times 16[(p\cdot k -p\cdot p^\prime)(p^\prime\cdot k -p\cdot p^\prime)+m_q^4]/(4(p\cdot k)(p^\prime \cdot k)) \nonumber \\
+16[p\cdot k (p^\prime \cdot k) -m^2p\cdot p^\prime + m^2(p\cdot k + p^\prime \cdot k) -m^4]/(4(p^\prime\cdot k)^2) \bigg\} {\rm .} \\ \nonumber
\end{eqnarray}
The factor of 1/2 in the front of this expression comes from the $\pi^0$ being a superposition of up and down quarks; it can be determined more carefully by examining the isospin 
matrices in the quark-meson model, but this will provide no great insight compared with the previous statement. Here and in the following work, we ignore the degeneracy factors related to the colors of 
the quarks, as they multiply all rates by the same factor and therefore will cancel out in ratios such as the quantum efficiency. 

Similarly, the production of a single charged pion and a photon $q\bar{q}\to \pi^\pm \gamma$ leads to the following matrix element squared:
\begin{eqnarray}
& & \sum |{\cal M}_{\gamma \pi^\pm}|^2 = \bar{v}^\prime(i\gamma_5)\frac{i(\slashed{p}-\slashed{k}+m)}{(p-k)^2-m^2}(2ie/3)\slashed{\epsilon}u\left(\bar{v}^\prime(i\gamma_5)\frac{i(\slashed{p}-\slashed{k}
+m)}{(p-k)^2-m^2}(2ie/3)\slashed{\epsilon}u\right)^* \nonumber \\ 
&+& 
\bar{v}^\prime(ie/3)\slashed{\epsilon}\frac{i(\slashed{k}-\slashed{p}^\prime+m)}{(k-p^\prime)^2-m^2}(i\gamma_5)u
\left(\bar{v}^\prime(ie/3)\slashed{\epsilon}\frac{i(\slashed{k}-\slashed{p}^\prime+m)}{(k-p^\prime)^2-m^2}(i\gamma_5)u \right)^*
\nonumber \\
&+& \bar{v}^\prime i\gamma_5 u \frac{i}{(p+p^\prime)^2-m_\pi^2}(+ie)(2(p+p^\prime)-k)^\mu \epsilon_\mu
\left(\bar{v}^\prime i\gamma_5 u \frac{i}{(p+p^\prime)^2-m_\pi^2}(+ie)(2(p+p^\prime)-k)^\mu \epsilon_\mu \right)^* \nonumber \\
&+& \bigg[ \bar{v}^\prime(i\gamma_5)\frac{i(\slashed{p}-\slashed{k}+m)}{(p-k)^2-m^2}(2ie/3)\slashed{\epsilon}u
\left(\bar{v}^\prime(ie/3)\slashed{\epsilon}\frac{i(\slashed{k}-\slashed{p}^\prime+m)}{(k-p^\prime)^2-m^2}(i\gamma_5)u \right)^* + c.c. \bigg] \nonumber \\
&+& \bigg[ \bar{v}^\prime(i\gamma_5)\frac{i(\slashed{p}-\slashed{k}+m)}{(p-k)^2-m^2}(2ie/3)\slashed{\epsilon}u
\left(\bar{v}^\prime i\gamma_5 u \frac{i}{(p+p^\prime)^2-m_\pi^2}(+ie)(2(p+p^\prime)-k)^\mu \epsilon_\mu \right)^* + c.c. \bigg] \nonumber \\
&+& \bigg[ \bar{v}^\prime(ie/3)\slashed{\epsilon}\frac{i(\slashed{k}-\slashed{p}^\prime+m)}{(k-p^\prime)^2-m^2}(i\gamma_5)u 
\left(\bar{v}^\prime i\gamma_5 u \frac{i}{(p+p^\prime)^2-m_\pi^2}(+ie)(2(p+p^\prime)-k)^\mu \epsilon_\mu \right)^* + c.c. \bigg] \nonumber
\end{eqnarray}
\begin{eqnarray}
&=& \left( \frac{4e^2g_{q\bar{q}\pi}^2/9}{(2p\cdot k)^2} +\frac{e^2g_{q\bar{q}\pi}^2/9}{(2p^\prime \cdot k)^2}\right)16\left[ (p^\prime \cdot k)(p \cdot k) +m^2_q(p\cdot k + p^\prime\cdot k)-m^2_q p^\prime
 \cdot p -m^4_q \right] \nonumber \\
&-& \frac{e^2g_{q\bar{q}\pi}}{(2m_q^2-m_\pi^2+2p\cdot p^\prime)^2} 16(p\cdot p^\prime + m^2_q)(2m^2_q+2p\cdot p^\prime-p\cdot k-p^\prime \cdot k) \nonumber \\
&-& 2 \frac{2e^2g_{q\bar{q}\pi}/9}{(2p\cdot k)(2p^\prime \cdot k)} 16\left((p\cdot k-p\cdot p^\prime)(p^\prime \cdot k -p\cdot p^\prime)+m^2_q p\cdot p^\prime \right) \nonumber \\
&+& 2 \frac{2e^2g_{q\bar{q}\pi}/3}{((2m^2_q-m^2_\pi+2p\cdot p^\prime)(2p\cdot k)}8(m^2_q+p\cdot p^\prime)(2m^2_q+2p\cdot p^\prime-p\cdot k -2 p^\prime \cdot k) \nonumber \\
&+& 2\frac{e^2g_{q\bar{q}\pi}/3}{(2m^2_q-m^2_\pi+2p\cdot p^\prime)(2p^\prime \cdot k)}8(m^2_q+p\cdot p^\prime)(2m^2_q+2p\cdot p^\prime -p^\prime\cdot k -2p\cdot k)
\end{eqnarray}

Finally, as emphasized earlier, the matrix elements squared for pion production without photons must be determined if we are to estimate quantum efficiency. We approximate 
\begin{equation}
\sum |{\cal M}_{2\pi^0}|^2 = \sum |{\cal M}_{\pi^0 \pi^+}|^2 = \sum |{\cal M}_{\pi^0 \pi^-|}|^2 {\rm ,} \nonumber
\end{equation}
which is exactly the approximation of isospin symmetry. The matrix element squared and summed over quark spins is
\begin{eqnarray}
\sum |{\cal M}_{2\pi^0}|^2 = 2g_{q\bar{q}\pi}^4\left[ \frac{(p\cdot k - p^\prime \cdot k)^2}{p\cdot k p^\prime \cdot k} + 1 + \frac{m^2(p\cdot p^\prime - m^2)}{p\cdot k p^\prime \cdot k} \right]{\rm .}
\end{eqnarray}

\subsection{Thermal rates of production at $T = 175\;{\rm MeV}$}
\label{sec:rates}

Thermal rates at $T = 175\; {\rm MeV}$ are used to determine the quantum efficiency of photon production as the plasma thermalizes. The rate integrated over quark and mesonic states is, in 
general,
\begin{eqnarray}
E_f\frac{d\Gamma_f}{dk^3_f} &=& \int \frac{d^3p_1}{(2\pi)^3(2E_1)}\frac{d^3p_2}{(2\pi)^3(2E_2)}\frac{d^3k_\pi}{(2\pi)^3(2E_{\pi})} f(E_1)f(E_2)|{\cal M}_{p_1p_2\to f\pi}|^2(1 + f(E_\pi)) \nonumber \\
& & \times (2\pi)^4\delta^4(p_1+p_2-k_f-k_\pi){\rm .} \nonumber
\end{eqnarray}
The factors of $2E$ come from the normalization of field operators used to define the states in the matrix elements, and, when written in combination with the integral measures, make Lorentz-invariant 
combinations. 

Integration over ${\bf k}_\pi$ and over one of the initial particles' azimuthal angle simplifies the integral into a 4-dimensional integral:
\begin{eqnarray}
E_f\frac{d^3\Gamma}{dk^3_f} &=& \int \frac{d^3k_\pi}{(2\pi)^32E_3}\frac{d^3p_1}{(2\pi)^32E_1}\frac{d^3p_2}{(2\pi)^32E_2}f_{FD}(E_1)f_{FD}(E_2) \sum |{\cal M}|^2
(2\pi)^4\delta^4(p_1+p_2-k_f-k_\pi) \nonumber \\
&=&  \frac{1}{8(2\pi)^6}\int \frac{d^3p_1}{E_1}\frac{d^3p_2}{E_2}f_{FD}(E_1)f_{FD}(E_2) \sum |{\cal M}|^2 \frac{2\pi}{E_3}\delta\left(E_1+E_2-E_f-E_\pi \right) 
\nonumber \\
&=& \frac{1}{8(2\pi)^4}\int \frac{p_1dp_1}{\sqrt{m^2_q+p^2_1}}\frac{p_2dp_2}{\sqrt{m^2_q+p^2_2}}d\theta_1 d\theta_2 d\phi f_{FD}(E_1)f_{FD}(E_2) \sum |{\cal M}|^2 \nonumber \\
& & \times \frac{\delta(\phi-\phi_r)+\delta(\phi+\phi_r)}{|\sin(\phi_r)|}{\rm .}
\label{thermal}
\end{eqnarray}
In the final line of Eq. \ref{thermal}, the angle $\phi$ is the difference in azimuthal angle between initial states 1 and 2, and $\phi_r$ is one of the solutions to 
\begin{eqnarray}
\cos(\phi_r) &=& \frac{2m^2_q + (E^2_f-m^2_\pi-|{\bf k}|^2) + 2(E_1E_2-E_1E_f-E_2E_f) + 2({\bf p}_1 \cdot {\bf k} + {\bf p}_2 \cdot {\bf k})}{2p_1p_2\sin(\theta_1)\sin(\theta_2)} \nonumber \\
& & + \cot(\theta_1) \cot(\theta_2) \nonumber
\end{eqnarray}

This formula will be used to calculate numerically the rates of both photon and pion production, near thermal freeze-out, in the quark-meson model. With what should these results be compared? The 
PHENIX collaboration has measured both the photon yields \cite{Adler:2005ig} and the $\pi^+$ yields \cite{PHENIXURL} in the 10-20\% centrality class. The plot of these yields is shown in Fig. \ref{fig:photonsVspiplus}.
Here lies the utility of using rates as opposed to lifetimes: if photons and pions were made exclusively by the processes in Section \ref{sec:quark-meson}, then the ratio of these two experimental yields 
would be equal to the ratio of the respective differential thermal rates. A fit to the ratio of the experimentally measured yields is shown in Fig. \ref{fig:ratio10to20}.

Using Eq. \ref{thermal}, the thermal rates can be easily calculated after making the proper substitutions for the matrix element squared. For photon production, this term is 
\begin{equation}
\sum |{\cal M}_{i\to f\gamma}|^2 = \sum |{\cal M}_{q\bar{q}\to \pi^0\gamma}|^2 + 2\sum|{\cal M}_{q\bar{q}\to \pi^+\gamma}|^2{\rm .} \nonumber
\end{equation}
The factor of two in the above equation comes from the matrix element squared with a $\pi^-$ in the final state is identical to the matrix element squared with a $\pi^+$ in the final state. For pion 
production, isospin symmetry is useful for simplifying the expression:
\begin{equation}
\sum |{\cal M}_{i\to f\pi^0}|^2 = 2\sum |{\cal M}_{q\bar{q}\to 2\pi^0}|^2 + \sum |{\cal M}_{q\bar{q}\to \pi^0\pi^+}|^2 \approx 4 \sum |{\cal M}_{q\bar{q}\to 2\pi^0}|^2{\rm .}\nonumber
\end{equation}
The factor of two in front of the first process comes simply from there being two neutral pions in the final state.

\begin{figure}[ht]
  \centering
  \includegraphics[trim=0cm 0cm 4cm 0cm, clip=true, width=12cm]{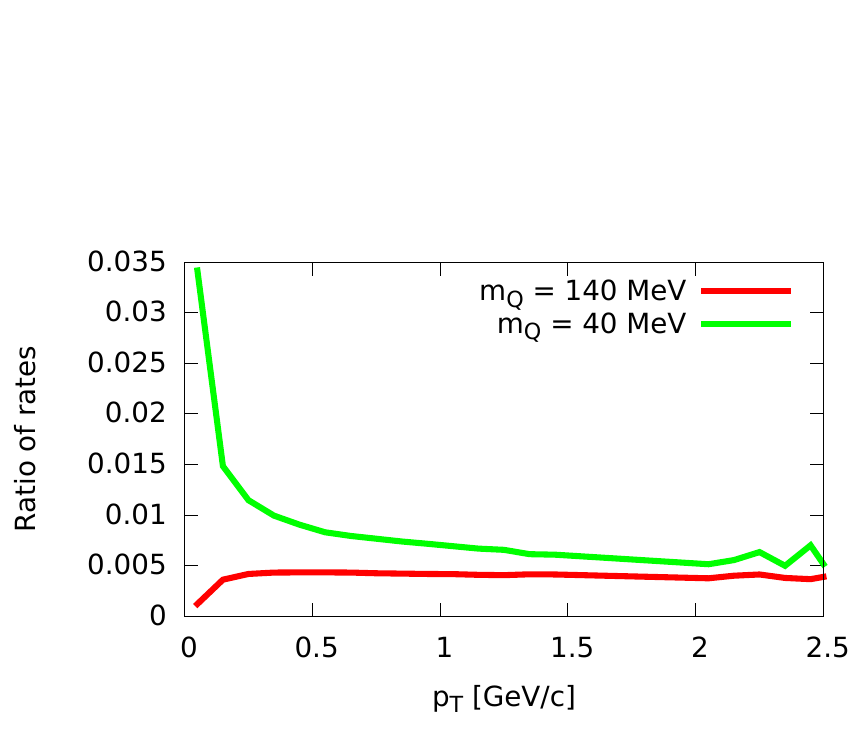}
  \caption{(Color online) the ratio of rates, for two different choices of the effective quark masses.}
  \label{fig:EdNd3k}
\end{figure}

The ratio of the thermal production rates is shown in Fig. \ref{fig:EdNd3k}. Here, the temperature is set to 175 GeV, every pion mass is set to $139.57\; {\rm GeV/c^2}$, and two dramatically different 
quark masses have been chosen, $140\; {\rm GeV/c^2}$ and $40\; {\rm GeV/c^2}$. Near the freeze-out temperature, the effective mass of quarks in any effective description of nuclear physics is expected to 
change rapidly, as the quarks are bad quasiparticles below the transition and good quasiparticles at sufficiently high temperatures. Both masses give similar results at high ${\bf k}$, but 
are different at low ${\bf k}$ where the radically different kinematical cuts have a significant effect.
We note that the simplest estimate of this ratio, $e^2/g^2_{q\bar{q}\pi} \approx 0.7\%$, is approximately one third of the experimentally determined ratio near $p_T=$ 2 GeV/c. 

Taking into account various effects will modify this calculation of the $\gamma/\pi$ ratio: 
first, the ratio would increase if the coexisting phase lasts long enough that mesons are created and destroyed numerous times. Second, if processes such as $q\bar{q}\rightarrow \rho\rightarrow\pi\pi$
were taken into account, the denominator would increase and the estimate for the ratio would decrease. Finally, the value of $g_{qq\pi}$ might be lower than the value of 3.63 used here, which was 
motivated by the Goldberger Treiman relation \cite{Bhaduri}. In fact, the value of $g_{q\bar{q}\pi}(T)$ found in \cite{Suzuki:1983ej} from fits to data for the decay of rho mesons to dielectrons 
is $2.97$, significantly lower than the estimate used in this paper. This would lead to a significant enhancement of the quantum efficiency of photon production, by a factor $(3.63/2.97)^2=1.5$.

\section{Enhanced production of excited states}
\label{sec:l>0}

All deconfined quarks must eventually disappear and form hadrons and other color singlets. At this transition to bound states, one might imagine a quark and an anti-quark beginning a spiral around 
each other. The quarks' momenta and separation determine the angular momentum and radial quantum number of the bound state. At a temperature of 175 MeV, the density of the plasma is such that a 
quark will on average form a color singlet with an antiquark at a distance of 1.3 fm. This is somewhat larger than a typical hadron. This suggests an enhanced production of meson-like bound states 
with $J>0$.

To estimate the production of these states, we now must go beyond the point-like descriptions of mesons in the previous section and use valence quark models for mesons. This will require 
finding solutions to the Dirac equation in spherical coordinates. A pedagogical review of these solutions is found in Appendix \ref{Dirac}. For the following discussion, we call the vector potential 
$V(r)$, the scalar potential $U(r)$, and we emphasize that the $a$-states have parity $(-1)^{j-1/2}$ while the $b$-states have parity $(-1)^{j+1/2}$. A good set of quantum numbers for 
these states contains the radial number $n$, angular momentum quantum number $j$, azimuthal quantum number $m$, and parity. The biggest weakness of the valence quark model used here is the lack of 
magnetism, which should be a very important effect for nearly massless quark and antiquarks.

We will treat the transition from quark to bound states as adiabatic: here, time-dependent potentials change slowly enough that the quantum numbers of a state are conserved. 
The adiabatic limit suffers from an important flaw: it ignores the collisions between quarks which may also change the angular momentum states. 
Rather than attempt to estimate this, we suggest that the yields of the hadrons described with the cocktail {\it be independently verified}, partly for the sake of understanding photon production.
We estimate the structure of the plasma near the transition: at the threshold of hadronization, we approximate the quarks as massless and free within some range, but also confined to form a color 
singlet with an antiquark in a spherical volume.
To confine a particle in a spherical volume of radius $R$, $U(r)$ is needed (the temporal component of a vector field does not confine massless particles). For massless Dirac particles confined in 
an infinite spherical well where $U(r)$ is zero below $R$ 
and infinite above $R$, the solutions to the radial wavefunctions are $A(r) = NJ_{j-1/2}(kr)$, $a(r) = NJ_{j+1/2}(kr)$ and $B(r) = NJ_{j+1/2}(kr)$, $b(r) = -NJ_{j-1/2}(kr)$. The momentum $k$ is 
determined by requiring the current normal to the spherical boundary of the well to be zero (interestingly, this is different from requiring $\psi({\bf x})$ to be zero):
$J_{j-1/2}(kr)-J_{j-1/2}(kr) = 0$ for a-states and $J_{j-1/2}(kr)+J_{j+1/2}(kr) = 0$ for b-states. The energy spectrum is listed in Bhaduri in units of $1/R$ \cite{Bhaduri}.
This relativistic generalization of the infinite square well is often called the ``bag model'', and is useful for describing some aspects of the hadronic spectrum \cite{Chodos:1974je}.
For the rest of the paper, we simply imagine hadronization as the shrinking of this bag from a large size to approximately the radius of a meson, $\approx$0.8 fm.

Using this, the energy spectrum of quarks confined to a spherical volume can be calculated as a function of $R$, the radius of the confining sphere, and thermal expectation values can be calculated 
using this spectrum. We work at $T = 175\;{\rm MeV}$. The average number of particles $\langle N(R) \rangle = \sum_i \frac{g_i\exp(-E_i(R)/T)}{1+\exp(-E_i(R)/T)}$ approaches 
$n_\infty \times \frac{4}{3}\pi R^3$, the number density in the infinite limit times the volume, fairly quickly; by $R=2\;{\rm fm}$, these numbers are nearly the same. We find the radius where there 
is on average 1 quark for each antiquark of the opposite color to be very nearly $1.3\;{\rm fm}$ at this temperature. In other words, the spatial structure of the plasma can be approximated with 
confining spheres of radius $R=1.3\;{\rm fm}$ containing on average 9 quarks and 9 antiquarks, which form 9 color singlets (ignoring baryons).

\begin{table}[h!]
  \begin{center}
    \caption{The energies of various eigenstates of the quarks confined to a sphere of radius 1.3 fm, and the probability of a quark to be found in that state at $T=175\;{\rm MeV}$.}
    \label{N_i}
    \begin{tabular}{c | c | c | c | c}
    A(B) & j & n & E [${\rm fm}^{-1}$] & $\langle N \rangle$ \\
      \hline
    A & 1/2 & 0 & 2.043 & 0.288 \\
    B & 1/2 & 0 & 3.812 & 0.070 \\
    A & 3/2 & 0 & 3.204 & 0.231 \\
    B & 3/2 & 0 & 5.123 & 0.046 \\
    A & 5/2 & 0 & 4.327 & 0.136 \\
    B & 5/2 & 0 & 6.371 & 0.024 \\
    A & 7/2 & 0 & 5.430 & 0.071 \\
    B & 7/2 & 0 & 7.581 & 0.012 \\
    A & 1/2 & 1 & 5.396 & 0.018 \\
    B & 1/2 & 1 & 7.002 & 0.005 \\
    A & 3/2 & 1 & 6.758 & 0.011 \\
    \end{tabular}
  \end{center}
\end{table}

We may now ask: what is the probability that these nearly free quarks and antiquarks form bound states with various quantum numbers? In the adiabatic transition, this is estimated by looking at 
the quantum numbers of the quarks and antiquarks before the transition. The thermal expectation value for a quark to be in a given state is given by
\begin{equation}
\langle N_i(R) \rangle = g_i\left(\frac{\exp(-E_i/T)}{1+\exp(-E_i/T)} \right)/\sum_j {\textstyle \frac{\exp(-E_j/T)}{1+\exp(-E_j/T)}}{\rm .}
\end{equation}
This is shown in Table \ref{N_i} for some of the lowest 
energy states and at $T=175\;{\rm MeV}$. The adiabatic transition has led to a significant enhancement of excited states compared with the thermal expectation values associated with the various mesons.
This strongly suggests that the hadronic cocktail used to determine the direct contribution to photon yields might be significantly underestimating the component coming from the 
decays of $a_n$ hadrons. 

Finally, the adiabatic limit is only one extreme limit for time-dependent perturbation theory. The other limit is the ``sudden approximation'', where free quarks and antiquarks are immediately subjected 
to a confining potential. This is a very interesting limit theoretically for relativistic wave equations: the vacuum of the field theory becomes nontrivial to define. A universal feature of all 
relativistic field theories with acceleration is the production of particle-antiparticle pairs, for a pedagogical review of this see \cite{Carroll}. We considered this extreme limit as well, for 
the case of massless quarks subjected suddenly to a confining potential; the results are summarized in Appendix \ref{sec:sudden}.

\section{Electromagnetic transitions of the excited states}
\label{sec:EM}

These excited states must ultimately decay. Electromagnetic transitions are possible, and will contribute to the production of photons at freeze-out. Ideally, one would estimate the enhancement of 
excited states at freeze-out, map the quantum numbers of these excited states to measured hadronic states, and use this result to modify the cocktail contribution to inclusive photon production. 
However, the lack of measurements of the electromagnetic decays of these states and in some cases, the masses of these states, makes such a calculation difficult at the moment. We end this work by 
estimating the rates for spontaneous emission, using the same models for mesons used in Section \ref{sec:l>0}.

The decay rate for a given transition, $\Gamma_{i\rightarrow f,{\bf k}}$, is found using perturbation theory,
\begin{eqnarray}
\Gamma_{i\rightarrow f,{\bf k}}&=&\frac{k}{3\pi}
\left|\langle f|\vec{\alpha}e^{i{\bf k}\cdot{r}}|i\rangle\right|^2.
\end{eqnarray}
Here $\vec{\alpha}$ are the $\alpha$ matrices used in the Dirac representation of the Dirac equation,
\begin{eqnarray}
\vec{\alpha}=\left(\begin{array}{cc}0&\vec{\sigma}\\ \vec{\sigma}&0\end{array}\right).
\end{eqnarray}
This equation was simplified by applying the dipole approximation, $e^{i{\bf k}\cdot{\bf r}}=1$, which unlike the case for atomic or nuclear transitions, is not a particularly good approximation for 
massless quarks. With this approximation decays are confined to final states with $j$ within one unit of the decaying state, and the new state must have opposite parity. If the dipole approximation 
were relaxed, the allowed matrix elements would be reduced by the phase factor, but transitions to other states would then be possible. It should also be emphasized that this picture ignores the fact 
that the initial and final states are complex many-body states and that the true matrix element might be significantly lower (as is represented by spectroscopic factors). Nonetheless this provides 
a starting point for understanding the potential impact of these decays.

Using the wavefunctions in Appendix \ref{Dirac}, the matrix elements can be significantly simplified. Summing over final-state polarizations $m_f$, 
\begin{eqnarray}\nonumber
\Gamma_{i\rightarrow f}&=&\frac{ke^2}{3\pi}
\left\{\begin{array}{ll}
\frac{2j_f+1}{j_i}\left( \int_0^\infty r^2dra_f(r)A_i(r)\right)^2,&j_f=j_i-1,~a\rightarrow a\\
\frac{2j_f+1}{j_i}\left( \int_0^\infty r^2drB_f(r)b_i(r)\right)^2,&j_f=j_i-1,~b\rightarrow b\\
\frac{1}{j_i(j_i+1)}\left(\int_0^\infty r^2dr [(j_i+1)A_f(r)b_i(r) -j_i a_f(r)B_i(r)]\right)^2,&j_f=j_i, b\rightarrow a\\
\frac{1}{j_i(j_i+1)}\left(\int_0^\infty r^2dr [(j_i+1)b_f(r)A_i(r) -j_i B_f(r)a_i(r)]\right)^2,& j_f=j_i, a\rightarrow b\\
\frac{2j_f+1}{j_f}\left( \int_0^\infty r^2dr A_f(r)a_i(r)\right)^2,& j_f=j_i+1, a\rightarrow a\\
\frac{2j_f+1}{j_f}\left( \int_0^\infty r^2dr b_f(r)B_i(r) \right)^2,& j_f=j_i+1, b\rightarrow b\\
\end{array}\right.\\
\end{eqnarray}
The charge, $e$ needs to be altered to fit the charge of the given quark that is undergoing the transition.

The rates for these transitions can now be determined with these approximations.
Table \ref{radiative} shows the numerical results for some of the transitions among the energy eigenstates when the potential is the Cornell potential. 
When $\Delta n = 0$, they are on the order of 0.001 ${\rm fm}^{-1}$.
Since the total rate of decay of each of these states is similar to 1 ${\rm fm}^{-1}$, the branching fraction into inclusive photons 
is roughly one tenth of a percent.

\begin{table}[h!]
  \begin{center}
    \caption{The rates for a representative sample of transitions from $(n_1, j_1)\to (n_2, j_2)$. The factor of $1/q^2$ represents the fractional charge of the quark or antiquark.}
    \label{radiative}
    \begin{tabular}{c | c | c | c | c}
    $j_1$ & $n_1$ & $j_2$ & $n_2$ & $\Gamma/q^2$ \\
      \hline
      3/2 & 0 & 1/2 & 0 & 0.003178 \\
      3/2 & 1 & 1/2 & 0 & 7.553e-05 \\
      3/2 & 1 & 1/2 & 1 & 0.001413 \\
      3/2 & 2 & 1/2 & 2 & 0.001010 \\
      3/2 & 3 & 1/2 & 3 & 0.0008215 \\
      3/2 & 4 & 1/2 & 4 & 0.0007087 \\
      3/2 & 4 & 7/2 & 3 & 2.259e-06 \\
      9/2 & 4 & 7/2 & 4 & 0.0007591 \\
      \end{tabular}
  \end{center}
\end{table}

If the population of states is given, one can then calculate the rate to emit photons of a given energy per excited state. Combined with an estimate of the density of states per unit volume, one can 
find $d\Gamma/dEd^4x$. The density of quarks would be $\approx 2.1$ per cubic fm for light free quarks, which suggests the density of proto-hadron is $\approx 1$ fm$^{-3}$. The photon yield per 
volume binned by energy can then be generated if one assumes that the rate roughly exists for a given time, which here we will assume that time is 3 fm/$c$. Finally, one can estimate the 
$\gamma/\pi^+$ ratio from entropy arguments. Lattice calculations \cite{lattice} show the entropy is $8/{\rm fm}^3$ when $T=175$ MeV. In the final state, there are approximately 4.5 units of entropy 
per particle and about 20\% of the particles are positive pions. This suggests that $1\; {\rm fm}^3$ of this matter should be responsible $\approx 0.3$ positive pions per cubic fm. 

\begin{figure}[ht]
  \centering
  \includegraphics[width=12cm]{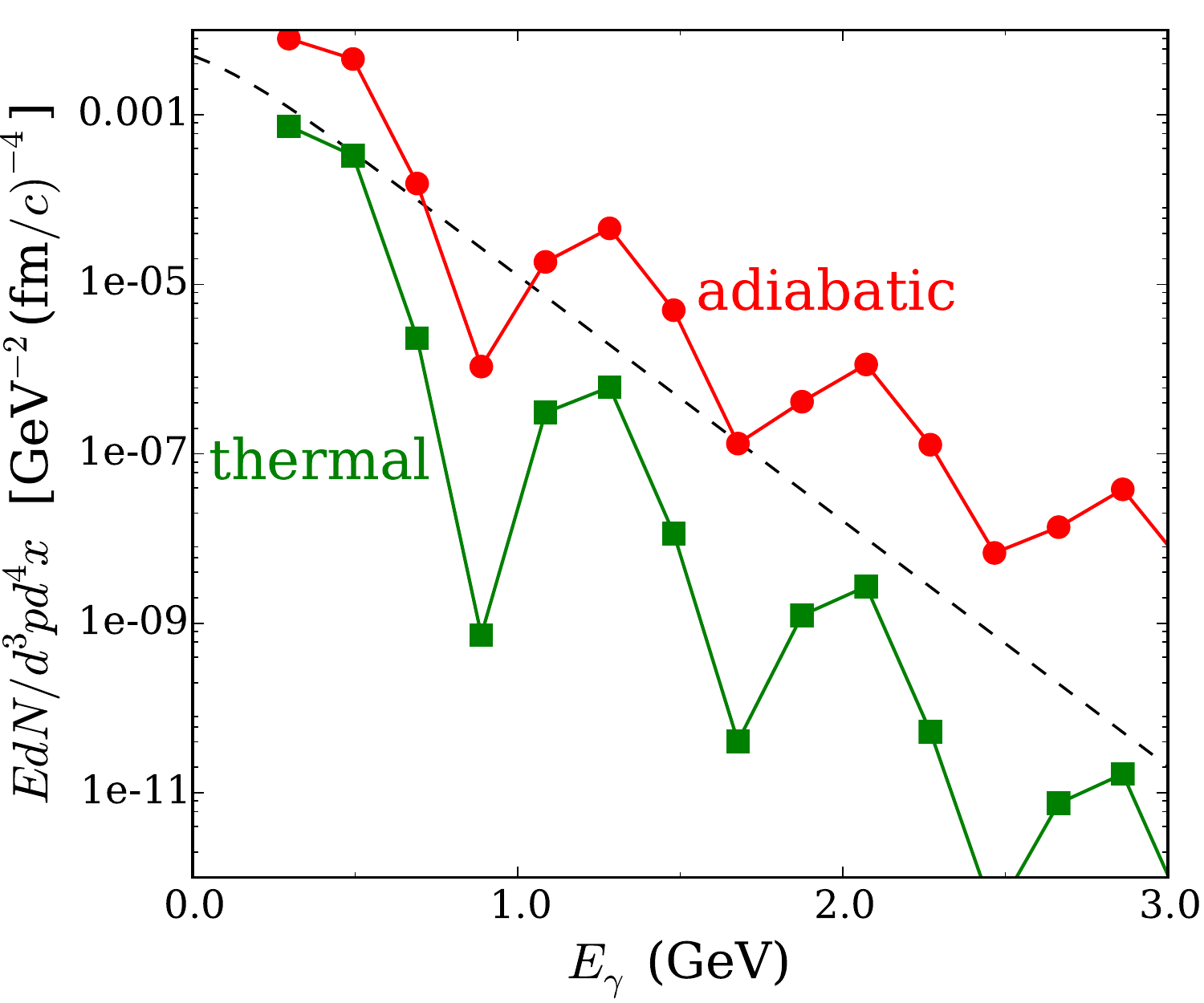}
  \caption{(Color online) the photon production rates for two different populations of excited bound states. The black dashed line shows is proportional to $\exp(-E/T)$, where $T=0.15\;{\rm MeV}$.}
  \label{fig:bothspectra}
\end{figure}

We consider an adiabatic transition where the radius of the confining bag $R$ slowly changes from 1.3 fm to 0.8 fm. 
In Figure \ref{fig:bothspectra}, the rates for photon production in two different scenarios are compared. In the curve marked ``adiabatic'', the populations of excited states are determined by 
thermal expectation values {\it before} the phase transition where the bag shrinks, when $R=1.3\;{\rm fm}$, with the energy levels determined by this large radius. 
In the curve marked ``thermal'', the populations of excited states are determined by the energy levels after the bag shrinks, when $R=0.8\;{\rm fm}$.

Two facts emerge from examining these curves. First of all, the adiabatic transition to hadrons has led to an enhancement of photon yields, exactly as described intuitively at the beginning of 
Section \ref{sec:l>0}. Second, there is some production of photons predicted in our valence quark model just coming from the excitations of hadrons, in a thermal, equilibrated hadron gas, as shown by 
the curve marked ``thermal''. Depending on the (temperature-dependent) masses of heavy mesons in hadron gas, it might be necessary to include their decay through photon emission when considering 
photon production at the late stage of the collision.
The ratio of the photon yields in Fig. \ref{fig:photonsVspiplus} to the rate per unit volume in
Fig. \ref{fig:bothspectra} at $p_T=2.5$ GeV is $\approx (10\; {\rm fm})^4$, suggesting that the decays of excited states enhanced at freeze-out might constitute a few
tenths of the production of ``direct photons'' near this momentum.

\section{Conclusions}
\label{sec:Conclusions}

Photons must be created during the recombination stage of heavy ion collisions. In this paper, we used a simplified model to show roughly the magnitude of this production. The yield falls short of 
dominating in any range of transverse momentum, however, it is nevertheless significant enough in the 2-3 GeV range so that further investigation is necessary. Significant uncertainty exists in our 
estimation, owing to the nature of the physics in the problem: we are estimating photon production in a thermal quark-meson model exactly where the parameters of the model should be changing rapidly, 
reflecting the change in the degrees of freedom from quarks to hadrons.

There is also theoretical evidence for enhancement of $J>0$ and radially excited mesonic states above thermal expectations at the point of hadronization. These will decay, also possibly by 
electromagnetic transitions which produce light at energies of approximately 1 GeV.

Finally, we exploited the valence quark model we used to describe mesons to examine the electromagnetic radiation of hadronic gas. The model had the strength of including all excited states, but 
the weakness shared by all simplified models for hadrons where the states do not exactly match the hadronic spectrum.

The photons produced at the point of recombination will have an elliptic flow similar to that of the hadrons detected in heavy ion collisions. While the yield of these photons is not enough immediately 
to explain the large measured $v_2$ of photons, it is large enough that it might be part of the eventual explanation of this effect.

We tried to be comprehensive by describing all the possible effects of the medium on this production of radiation. Significant areas for improvement 
of these calculations exist: first, the timescales over which quarks form bound states need to be examined carefully. Second, the models for mesons should be improved, perhaps with the inclusion of 
chromomagnetism. Next, because these calculations have implications not just for photons but also for the yields of heavy mesons, the calculations in Section \ref{sec:l>0} should be checked against 
and perhaps superseded simply by experimental measurements of the production of these states in heavy ion collisions.
Finally, the radiated photon's momentum is large compared with the binding energy of mesons, making the inclusion of electromagnetic form factors an important and relatively easy next step.

\section{Acknowledgments}

We especially thank Sarah Campbell and Richard Petti for helpful discussions.
This work was supported by the Department of Energy Office of Science through grant number DE-FG02-03ER41259.

\appendix

\section{The Dirac equation in spherical coordinates}
\label{Dirac}

The solutions to the Dirac equation in spherically symmetric potentials have been worked out pedagogically in a number of references \cite{Bhaduri, Abers}. It is possible to make simultaneous 
eigenstates of energy, $J^2$, $J_z$, and parity. 
In the Dirac representation, the four-component wave functions have the form
\begin{equation}
\psi_{a,n,j}({\bf r}) = \begin{pmatrix} A_{n,j}(r) {\cal Y}_{j-\frac{1}{2}}^{j,m} \\ -ia_{n,j}(r) {\cal Y}_{j+\frac{1}{2}}^{j,m} \end{pmatrix}
\label{centrala}
\end{equation}
where $\psi_{a,n,j}({\bf r})$ has parity $(-1)^{j-\frac{1}{2}}$, and 
\begin{equation}
\psi_{b,n,j}({\bf r}) = \begin{pmatrix} B_{n,j}(r) {\cal Y}_{j+\frac{1}{2}}^{j,m} \\ -ib_{n,j}(r) {\cal Y}_{j-\frac{1}{2}}^{j,m} \end{pmatrix}
\label{centralb}
\end{equation}
where $\psi_{b,J}({\bf r})$ has parity $(-1)^{j+\frac{1}{2}}$, and ${\cal Y}_\ell^J$ are two-component spinors,
\begin{equation}
{\cal Y}_\ell^{l\pm \frac{1}{2},m} \equiv
 \frac{1}{\sqrt{2l+1}} \begin{pmatrix} \sqrt{l\pm m + 1/2} Y _\ell^{m-\frac{1}{2}} \\ \pm \sqrt{l\mp m + 1/2} Y _\ell^{m+\frac{1}{2}}  \end{pmatrix} \nonumber\\
\end{equation}
As usual, the eigenvalue of $\hat{J}^2$ is $j(j+1)\hbar^2$ and of $\hat{J}_z$ is $m\hbar$. However, the orbital and intrinsic angular momentum quantum numbers cannot form a set of commuting observables, 
not even in the case of Dirac particles without electric charge, which is not the case for the Dirac equation's non-relativistic limit, the Schr\"odinger equation.

The radial functions $A_{n,j}(r)$ and $a_{n,j}(r)$ are solutions to the equations
\begin{eqnarray}
(E-V(r)-m-U(r))A_{n,j}(r) & = & a_{n,j}^\prime(r) + \frac{(j+3/2)a_{n,j}(r)}{r}{\rm ,}\\
(E-V(r)+m+U(r))a_{n,j}(r) & = & -A_{n,j}^\prime(r) + \frac{(j-1/2)A_{n,j}(r)}{r}{\rm .} \nonumber\\
\end{eqnarray}
The equations for $B_{n,j}(r)$ and $b_{n,j}(r)$ yield solutions with the same $n,\; j,$ and $m$ quantum numbers but opposite parity:
\begin{eqnarray}
(E-V(r)-m-U(r))B_{n,j}(r) & = & b_{n,j}^\prime(r) - \frac{(j-1/2)b_{n,j}(r)}{r}{\rm ,} \nonumber \\
(E-V(r)+m+U(r))b_{n,j}(r) & = & -B_{n,j}^\prime(r) - \frac{(j+3/2)B_{n,j}(r)}{r}{\rm .} \nonumber
\end{eqnarray}
Here, there are two possible spherically symmetric potentials: the scalar potential $U(r)$, which acts very much like a position dependent mass for the particles and antiparticles, and a vector potential 
$V(r)$, which is one component of the potential coming from a gauge interaction. 

For a particle with energy $E$, the antiparticle solution of energy $-E$ is found by solving the equations with the sign of the vector potential $V(r)$ reversed but without changing the sign of the 
scalar potential $U(r)$ nor the mass.

In the bag model for hadrons, quarks are confined to spherical volumes where the probability of finding a quark outside of some radius $R$ vanishes. Some work with the equations above shows that this 
is impossible simultaneously for particle and antiparticle solutions using a vector potential, but can be done with the scalar potential $U(r)$ (note how this matches the intuition of the scalar 
potential being a position-dependent mass, which can make both particles and antiparticles heavy, while Coulomb potentials affect particles with different charges differently).
For the bag model used in this work, the scalar potential is infinite for $r>R$, zero inside, and there is no vector potential. 
At the boundary, the solutions obey the boundary condition
\begin{equation}
i\gamma^\mu n_\mu\psi(R)=\psi(R),
\end{equation}
which comes from forcing the Dirac field's probability current $\bar{\psi}\gamma^\mu \psi$ to be zero at the boundary.
Note how the lack of a derivative term in the probability current for a Dirac particle potnentially leads to discontinuities at boundaries. 

Given that there is zero potential inside the solutions are given in terms of spherical Bessel functions,
\begin{eqnarray}
A_{n,j}(r)&=&J_{j-1/2}(E_nr),~a_{n,j}(r)=J_{j+1/2}(E_nr),\\
\nonumber
B_{n,j}(r)&=&-J_{j+1/2}(E_nr),~b_{n,j}(r)=J_{j-1/2}(E_nr).
\end{eqnarray}
For each $J$ and parity ($A,a$ solutions vs. $B,b$ solutions) there are numerous values of $E_n$ that satisfy the boundary conditions; they are indexed by $n$, which increases by one with each additional 
node. We emphasize one more time that these quantum numbers vary from the usual non-relativistic problem in that the orbital angular momentum is not a good quantum number, and 
the upper and lower components of the wave function behave as $\ell=J-1/2$ or $\ell=J+1/2$ for the $A,a$ solutions, and as $\ell=J+1/2$ or $\ell=J-1/2$ for the $B,b$ solutions.

\section{The ``sudden approximation'' and particle production in quantum field theories with time-dependent potentials}
\label{sec:sudden}

Hadronization is, in the most abstract sense, a problem of time-dependent potentials: deconfined quark states are subjected to a confining potential, leading to the production of hadrons and in this
paper, photons. The formalism most familiar to physicists for dealing with time-dependent potentials is the Dyson series in time-dependent perturbation theory. However, when starting work on a 
given problem, consideration of the extreme limits is helpful and occasionally, all that is needed.

In the main narrative of this paper, we considered hadronization to be an adiabatic process: the potential changes from negligible to confining slowly enough so that the {\it quantum numbers} of the 
various quark states are conserved. The opposite limit is the ``sudden approximation'', where the potential turns on instantaneously. In this limit, the {\it wavefunction} is conserved (the intuition 
is that not enough time has passed for the probability distribution to change). In non-relativistic quantum mechanics, this approximation is very useful: one simply takes the wavefunction right before 
the potential turns on, determines the expression for this wavefunction in terms of the {\it new} eigenfunctions of the system once the potential has turned on, and one now knows the dynamics of the 
system into the future. In relativistic field theories, the situation is richer physically, as we will now demonstrate.

We examined the particle production in the sudden approximation using central potentials which phenomenologically describe quarks in mesons.
The massless Dirac equation was solved numerically for the lowest energy $(-1)^{j-\frac{1}{2}}$-parity eigenstates when 
$j=\frac{1}{2}{\rm ,}\; \frac{3}{2}{\rm ,}\; \frac{5}{2}{\rm ,}\; \frac{7}{2}{\rm ,}$ and $\frac{9}{2}$. To be physically relevant to quarks forming hadrons, the central potentials were set as
$V(r)=-0.383/r$ and $U(r) = 5.73r/{\rm fm}^2$; the potential $U(r)$ confines all solutions to this equation while $V(r)$ is the Coulomb potential appropriate for weakly coupled gauge theories. 
These are the solutions appropriate for quarks suddenly forced into bound states by a rapid transition to low temperatures. Meanwhile, massless quarks above deconfinement have, in the Dirac 
representation of the gamma matrices, the normalized solution for a left-handed particle (not antiparticle) moving freely in the z-direction:
\begin{equation}
\psi_k(z,t) = \frac{1}{\sqrt{2}(2\pi)^{3/2}}\exp(-i|k|t+ikz)\begin{pmatrix} 1 \\ 0 \\ 1 \\ 0 \end{pmatrix}{\rm .}
\label{plane}
\end{equation}

If the potentials suddenly change from zero to the Coulombic and confining potential terms described above, one encounters a problem common when working with quantum fields in curved spacetime: 
it becomes impossible to choose a vacuum which is ``empty'' both when the potential is zero and when it is the confining Cornell potential. What was the vacuum state for free massless Dirac particles 
is now is an excited state with particles and antiparticles described by the $\psi_a$ and $\psi_b$ wavefunctions. Very abstractly, where the subscript ``1'' indicates the free theory and the 
subscript ``2'' the confined theory, the expectation for the number of particles of type ``2'' after the vacuum of type ``1'' is subjected to the confining potential is given by
\begin{equation}
\langle 0|_1 \hat{b}^{i\dagger}_2 b^i_2 |0\rangle_1 = \sum_{j,k} \langle 0|_1 (\alpha_{ij}b^j_1 + \beta_{ij}b^j_1)^\dagger (\alpha_{ik}b^k_1 + \beta_{ij}b^k_1) | 0 \rangle_1
= \sum_j \beta_{ij}\beta^\dagger_{ji}{\rm ,} \nonumber
\end{equation}
where $\alpha_{ij}$ and $\beta_{ij}$ are the Bogoliubov coefficients, the index $j$ represents labels for both momentum and spin and the index $i$ represents the quantum numbers of the $a$ and $b$ 
states, and where no sum over $i$ is represented by the repeated index. When the initial state is not the vacuum but instead contains a free Dirac particle, the expectation for the number of particles 
becomes
\begin{equation}
\langle 0|_1 b^l_1(k)  \hat{b}^{i\dagger}_2 b^i_2 b^{l\dagger}_1(k) |0\rangle_1 = |\beta_{il}|^2 + \sum_j \beta_{ij}\beta^\dagger_{ji}{\rm .} \nonumber
\end{equation}

In non-relativistic quantum mechanics (where there is no particle production by potentials breaking Poincare symmetry), the quantity $|\beta_{il}|^2$ would simply be the probability that a plane wave 
would be measured in the energy eigenstate described by the $\psi_{a(b)}$ wavefunctions:
\begin{equation}
P_{k->n,J,M,a(b)} = \left| \int d^3x \psi_k^\dagger(z)\psi^{n,J,M}_{a(b)}({\bf x})\right|^2 {\rm .} 
\label{overlap}
\end{equation}
More explicitly, for the $a$ states, 
\begin{eqnarray}
P_{k->n,J,M,a} &=& \bigg| \int_0^\infty r^2 dr \int_0^\pi \sin(\theta)d\theta \int_0^{2\pi} d\phi \frac{1}{\sqrt{2}}\frac{1}{(2\pi)^{3/2}}\exp(-ikz)
\bigg[ \sqrt{\frac{J+M}{2j}}A(r)Y_{J-1/2}^{M-1/2}(\theta,\phi) \nonumber \\
& & -i\sqrt{\frac{J-M+1}{2(J+1)}}a(r)Y_{J+1/2}^{M-1/2}(\theta,\phi)  \bigg] \bigg|^2 {\rm .} \nonumber
\end{eqnarray}
Here, the $z$-direction was chosen to coincide with the plane wave's momentum ${\bf k}$. This probability should then be 
convolved with the thermal distribution of quarks at $T=175\;{\rm MeV}$ and summed over $m$ states:
\begin{equation}
\Gamma_{n,J} = \int d^3p \exp(-p/T) \sum_m \left|\int d^3x \psi^\dagger_k({\bf x})\psi_a^{n,J,M}({\bf x})\right|^2{\rm .}
\label{j>0rate} 
\end{equation}

\begin{figure}[ht]
  \centering
  \includegraphics[trim=0cm 0cm 0cm 0cm, clip=true, width=11cm]{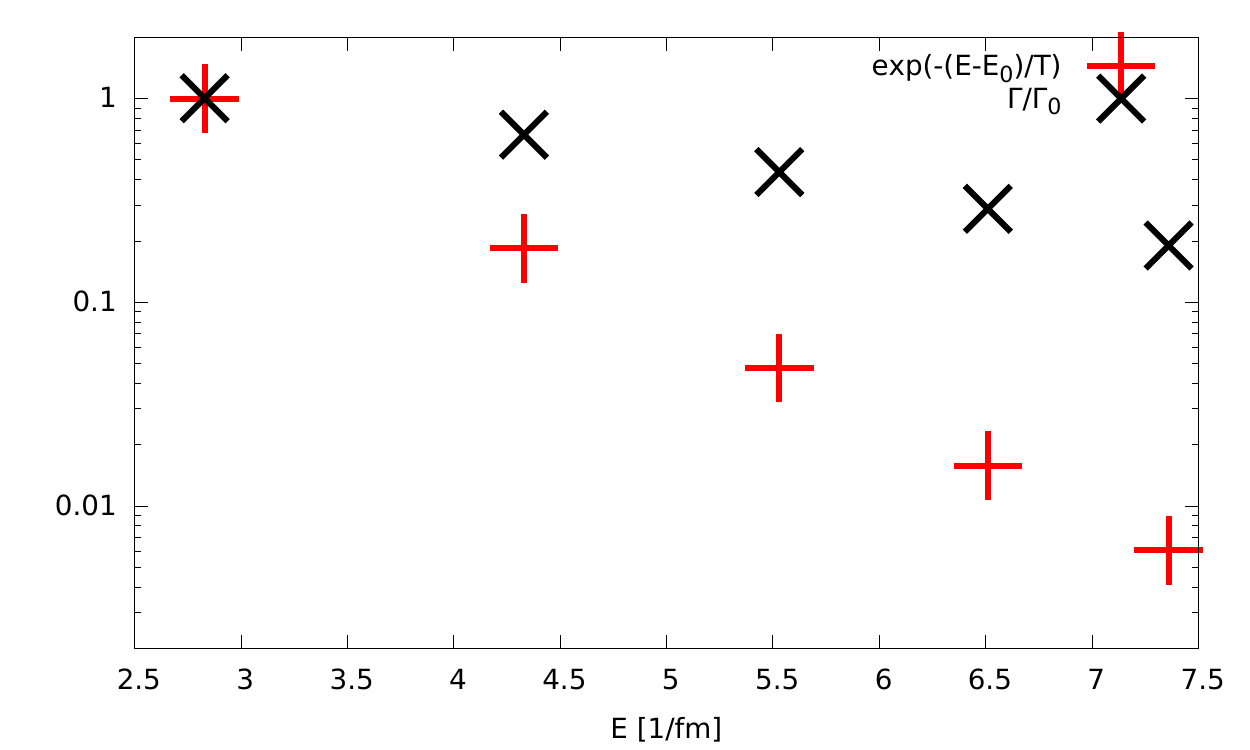}
  \caption{(Color online) The ratio $\exp(-E_j/T)/\exp(-E_{1/2}/T)$ of Boltzmann factors for the lowest energy $a$ state of a given $j$, compared with the ratio of thermal rates of production of 
  these states.}
  \label{fig:overlap}
\end{figure}

In Fig. \ref{fig:overlap}, the ratio of the thermal average of the probability of a massless quark with momentum $2\;{\rm fm}^{-1}$ to the probability of being found the in the $j=1/2$ lowest a state 
is compared with the ratio of the Boltzmann factor at $T=175\;{\rm GeV}$ to the Boltzmann factor of this ground state. Comparing these ratios, instead of just comparing the quantum-mechanical and 
thermal probabilities, spares us from having to calculate the partition function for a particle in these potentials. The ratios by definition agree for the ground state, while the $J>1/2$ states 
show significant enhancement above the thermal value.

\end{document}